\documentclass[10pt]{article}
\usepackage{cite}
\usepackage{amssymb}
\usepackage{pifont}\usepackage{amsmath}
\usepackage{subfigure}
\usepackage{graphicx}
\usepackage{pstricks}
\usepackage{pst-plot}
\usepackage{pst-slpe}
\usepackage{latexsym}
\allowdisplaybreaks[4]

\def\PREP(#1,#2,#3){Phys.\ Rep. \issue(#1,#2,#3)}
\def\EPJC(#1,#2,#3){Eur.\ Phys.\ J.\ C \issue(#1,#2,#3)}

\def\myblue#1{\bf\textcolor{blue}{\textsf{#1}}}

\newcommand{\beq}{\begin{equation}}
\newcommand{\eeq}{\end{equation}}
\newcommand{\bea}{\begin{eqnarray}}
\newcommand{\eea}{\end{eqnarray}}
\newcommand{\bfig}{\begin{figure}}
\newcommand{\efig}{\end{figure}}

\newcommand{\comment}[1]{}
\begin{document}

\begin{center}

{\Large \bf Soft supersymmetry breaking with tiny cosmological constant in flux compactified
$\mathcal{N}=1$ Supergravity
}
  \vglue 0.5cm
Debottam Das$^{(a,b)}$\footnote{debottam.das@th.u-psud.fr}, 
Joydip Mitra$^{(b,c)}$\footnote{tpjm@iacs.res.in},
Sudipto Paul Chowdhury$^{(b)}$\footnote{tpspc@iacs.res.in} and 
Soumitra SenGupta$^{(b)}$\footnote{tpssg@iacs.res.in}
    \vglue 0.2cm
{\em $^{(a)}$ Laboratoire de Physique Th\'eorique, UMR 8627, Universit\'e de Paris-Sud 11 \& CNRS, B\^atiment 210, 91405 Orsay Cedex,
France
\\}
{\em $^{(b)}$Department of Theoretical Physics \&
Centre for Theoretical Sciences, Indian Association
for the Cultivation of Science, Raja S.C. Mullick Road, Kolkata 700 032, India\\} 
{\em $^{(c)}$ Scottish Church College,1\&3 Urquhart Square, Kolkata 700 006, India\\}
    
\end{center}
\vspace{.2cm}


\begin{abstract} 
Using the flux compactification scenario in a generic supergravity model we propose a set of conditions which will  
generate de-Sitter or
anti de-Sitter vacua for appropriate choices of the parameters in superpotential. It is shown that a  
mass spectrum consistent with softly broken TeV scale supersymmetry in a minimal supersymmetric standard model 
at the observable sector 
can be obtained along with a tiny cosmological constant when the K\"ahler and superpotential of the hidden sector satisfy
a set of general constraints.
Constructing a specific model which satisfy the above constraints,
it is demonstrated that all the hidden sector fields have vacuum expectation values close to Planck scale and
the resulting low energy potential does not have any flat direction.
\end{abstract}

\section{Introduction}
Supersymmetry(susy) 
has emerged as one of the leading candidates for unraveling the physics beyond 
the Standard Model(SM). The most challenging and longstanding  
questions  in the context of supersymmetric theories are related to the mechanism of supersymmetry breaking.
The need for susy breaking originates from the fact that no superparticle has yet been observed in nature. 
The fine tuning/gauge hierarchy problem points towards the requirement of the susy breaking scale being close
to TeV so that the Higgs mass does not receive radiative corrections beyond TeV. 
As a result TeV scale supersymmetric theories have
drawn special attentions in view of the forthcoming results in the LHC. 
A possible origin of these symmetry breaking terms  
can be understood by embedding MSSM in 
a supergravity theory (SUGRA)\cite{nilles,gravmed2, grav-app}. Supergravity theories drew further attention when they were
shown to emerge as low energy effective theories of underlying string theories 
which can take care of the renormalizability problems
persistent in the supergravity models.  

In a plethora of supergravity models supersymmetry is assumed to be
spontaneously broken at a high energies ($\sim 10^{11}~\rm GeV$) via the vacuum expectation value(VEV) of the $F$ or $D$ 
terms \cite{soni, ibanez, barbieri, dudas, burgess1}.
Some realistic models of spontaneous susy breaking requires two different sets of superfields, one of which participate 
in Standard Model gauge interactions and the other consists of singlets under the SM gauge group. The singlet fields usually
have very large VEV and constitute the hidden sector of the theory.
Such a sector has it's natural origin in the framework of string theory. 
In a formalism where the matter fields of the MSSM constituting the observable sector
are assumed to couple with the hidden sector singlets via gravitational interactions, spontaneous susy breaking 
in the hidden sector results into generation of  soft-susy breaking terms in the observable sector\cite{soni,ibanez}.
If the local susy breaking is expected at a scale $\Lambda$, the soft terms would have a scale of 
order $\Lambda^2/M_p$, where $M_p$ is the Planck mass.
Considering $\Lambda \sim 10^{11}~\rm GeV$, soft supersymmetry breaking terms $\sim$ TeV can be produced at the observable
sector. 
Supergravity
models however are often encountered with the existence of flat directions in the hidden sector. 
It is therefore necessary to lift the flat directions by appropriate mechanism. 
  
While normally SUSY breaking leads to appearance of non-vanishing vacuum energy,
some of the of the SUGRA models may have broken SUSY with vanishing vacuum energy. 
The experimental data indeed 
indicates a near-flatness of the present universe \cite{WMAP3} with a vanishingly small vacuum energy. 
Simple model like no-scale supergravity scenario can give  
plausible explanation for nearly vanishing vacuum energy. However these models 
have limitations to produce viable susy phenomenology at low energies($\sim$TeV)
\cite{noscale1,noscale2}. 
Therefore the search for a near-Minkowski/de-Sitter vacua has attracted 
considerable interest in recent years. In this regard various supergravity and 
string motivated models generating de-Sitter vacua at the 
TeV scale have been 
proposed \cite{KKLT, Maldacena, choi, vempati, susskind, KKLT2, Copeland, 
Douglas, Blumenhagen, Burgess, Douglas2, Bertolini, Mambrini,Yoshioka,Watson,Misra}. 
Substantial progress have been made in this direction by a model 
developed in 2003\cite{KKLT} in the context of string theory known as the KKLT model. This was followed by a 
plethora of 
propositions \cite{choi, KKLT2, Copeland, Blumenhagen, ArkhaniHamed, 
Burgess,Douglas2, Bertolini, Mambrini}which revolve around building 
an effective description of low energy physics through a superpotential 
generated by background fluxes and a convenient choice of 
K\"ahler potential motivated by a suitable compactification scheme arising 
in String/M-Theory.\\
In this backdrop we address the following question in the present article :     
\begin{itemize}
\item What are the general set of criteria that the Kahler and superpotential of a supergravity model have
to satisfy to produce a viable phenomenology at the Tev scale so that there is no flat direction in the model and the
effective vacuum energy at the low energy is tiny and positive which is compatible to the present day observation?
\end{itemize}
Adapting the technique proposed in \cite{KKLT} 
an additional moduli field is incorporated in both the superpotential and the K\"ahler potential. 
As can be seen latter,
this extra moduli field is crucial in producing small vacuum energy, once 
susy is broken at the hidden sector.
A small vacuum energy at the minimum and viable susy breaking terms
at the observable sector restrict the choice of 
the K\"ahler and the superpotential.  
The order of magnitude of the gravitino mass 
and the soft susy breaking parameters namely the soft scalar mass, 
and the coefficient of the trilinear and the bilinear terms are estimated. 
This set-up is later illustrated with a toy model.

In section (1.1) of this article a brief review of \cite{KKLT} is presented. 
In section (2) an extension of the same has been suggested. 
Section (3) provides
the extremization condition for the scalar potential. 
Section (4) deals with the estimation of soft susy breaking parameters. 
In section (5) a possible model is proposed in support of our analysis.
Section (6) is devoted to summarizing the work.

\subsection{KKLT set-up : a quick revisit}
In this section a brief review of some parts of  the KKLT proposal relevant to the present work is presented. 
We begin with a quick revisit to the basic frame 
work of $\mathcal{N} = 1$ supergravity.
As mentioned in the introduction, the complete Lagrangian 
(up to two derivatives) in $\mathcal{N}=1$ supergravity is specified by chiral 
superfields $\phi_M$ of the theory where the index $M$ runs over the entire set of chiral superfields, the 
analytic gauge-kinetic function $f_{\alpha}(\phi_M)$, and the real 
gauge-invariant K\"ahler function  $G( \phi_M,{\phi_M}^*)$. 
The dimensionless function $G$ is defined 
in terms of K\"ahler potential $K( \phi_M,{\phi_M}^*)$  and 
superpotential $W(\phi_M)$ as,

\begin{equation}
\label{kfunc}
G( \phi_M,{\phi_M}^*)=\frac{K( \phi_M,{\phi_M}^*)}{M_p ^2}
+\log \mid \frac{W(\phi_M)}{M_p ^3} \mid^2{\myblue{,}}
\end{equation}
\noindent
where $M_p \sim 10^{19}~\rm GeV$ denotes the Planck scale.
The tree level(F-part) supergravity scalar potential in the Einstein frame 
is given by

\begin{equation}
\label{pot1}
V(\phi_M,{\phi_M}^*)=e^{K/M_p^2}\left(K^{I  \bar {J}} D_I W D_{\bar{J}} 
W^*-\frac{3}{M_p^2} \mid W(\phi_M) \mid^2 \right).
\end{equation}
\noindent
Here $D_I W=\partial_I W + \frac{K_I}{M_p^2} W$ is the K\"ahler covariant 
derivative of the superpotential and $K_{I {\bar J }}=
{\partial}_{\bar J}{\partial }_I K$ is the K\"ahler metric.
Now, using 
$F^M=M^3_p e^{G/2}K^{M \bar{P}} G_{\bar{P}}$ 
the scalar potential  can be recast as

\begin{equation}
\label{pot2}
V(\phi_M,{\phi_M}^*)=\left(\bar{F}^{\bar{N}} K_{\bar{N} M} F^M-3 M_p^4 
e^{G}\right).
\end{equation}
\noindent
In type-IIB String theory de-Sitter solutions are forbidden in the leading order of $ \alpha' $ 
(the string tension) and $g_s$ (the string coupling) by a 
no-go theorem \cite{Maldacena}. 
So a natural way of finding a de-Sitter solution in string theory 
is to include corrections 
from higher order in $\alpha'$ and $g_s$ to the no-scale structure. 
In \cite{KKLT} 
this objective is achieved in a two fold way :
 
\begin{enumerate}

\item 
In the presence of a non-zero flux the Calabi-Yau moduli combines 
suitably to form a superpotential of the form 

\begin{equation}
\label{Superpots}
 W = \int_M G_3 \wedge \Omega = \int_M (F_3 - \tau H_3) \Omega
\end{equation}

where $H_3$ and $F_3$ are the NS-NS and R-R three-form fluxes and $\tau$ is the axion-dilaton  of type-IIB string theory. 
The K\"ahler structure of the internal manifold (Calabi-Yau four-fold) is encoded in a K\"ahler potential whose tree-level form is 

\begin{equation}
\label{Kahlers}
{ K} = M^2_p(-3 ln[- i (\rho - \overline{\rho})] - ln[- i 
(\tau - \overline{\tau})] - ln [\int_M \Omega \wedge \overline{\Omega}]), 
\end{equation}

where $\rho$ is the single volume modulus. 

This structure however does not admit any de-Sitter 
solution and therefore 
calls for inclusion of corrections to 
the leading order.  
The relevant correction terms can be attributed to two different sources. 
One of these, being instanton correction which at large volume of the internal four-fold yields 
an additional term to the Superpotential 

\begin{equation}
\label{correct1}
{ W_{inst}} = P(z_i) e^{2 \pi i \rho}. 
\end{equation}

While $P(z_i)$ is the holomorphic one-loop determinant, the leading order 
in the exponential originates from the action of 
Euclidean $D3$-branes 
wrapping on a four-cycle. 
With all other moduli fixed, they can be integrated out to produce an effective 
superpotential for $\rho$ alone. 

The origin of the other correction is described as follows. 
In type II-B string theory compactified on a 
Calabi-Yau fourfold a stack of $N$ number of D7 branes wrapping 
on 4-cycles in the compact manifold gives rise to non-abelian gauge groups. 
The effective low-energy $\mathcal{N}=1$ supersymmetric theory   
undergoes gluino condensation resulting into a  
superpotential, 

\begin{equation}
\label{superpotcorrec}
 W_{gauge}=\Lambda_{N}^3=A e^{\frac{2 \pi i \rho}{N}}.
\end{equation}

Here  $\Lambda_{N}$ is the dynamical scale of gauge theory. The 
coefficient $A$ can
be determined by the energy scale below which Supersymmetric QCD is valid.
Plugging $\rho = i T$ in the superpotential along with a logarithmic 
$T$ dependence in the  K\"ahler potential we arrive at,
\begin{equation}
\label{superpot4} 
 W=W_0 + A e^{- a \tilde{T}}~,~~~  K=-3 M_p^2 \log(\tilde{T} + 
\tilde{T}^*), 
\end{equation}

where $\tilde T=\frac{T}{M_p}$.\\
In a supersymmetric vacuum $D_T {W} = 0$. The axion field 
in $\rho$ is set to zero for the sake of 
simplification. $\{A, a, W_0\}$ is taken to be real and $W_0$ to be negative. 
The condition for susy preserving 
vacua corresponds to the minimum of the scalar potential determined by the 
VEV of $T$.

\begin{equation}
D_T  W=0 \Longrightarrow W_0=-A e^{-a \tilde{t_0}}(1 + \frac{2}{3} a 
\tilde{t_0})\label{condition}.
\end{equation}

Here $\tilde t_0$ denotes the VEV of $\tilde T$. This makes the scalar 
potential (\ref{pot1}) negative at the minimum and can be given by,

\begin{equation}
\label{adspot}
V_{AdS}=-(3 e^{ K/M_p^2} {W}^2/M_p^2)_{AdS}={\frac{1}{M_p^2}}\frac{-a^2 A^2 e^{-2 a \tilde{t_0}}}
{6 \tilde{t_0}}
\end{equation}

Thus a supersymmetry preserving AdS minimum has been generated. 
Fixing the values of $\{A,a, W_0\}$, one can estimate the VEV of the 
moduli $T$ via (\ref{condition}) which in turn determines the $V_{AdS}$
through (\ref{adspot}).

\item
The next important step is to add a de-Sitter lifting term to
the scalar potential. The principal issue is to allow for generation of some extra energy from the 
flux background which shows up in the scalar potential as  an extra term namely $D/T^3$. 
This is achieved by incorporating anti $D3$-branes ($\overline{D3}$-branes) in the picture. The coefficient 
D depends on the number of $\overline{D3}$-branes introduced. 
The introduction of $\overline{D3}$-brane 
breaks supersymmetry and the results in a positive definite 
vacuum energy by compensating the AdS value of the scalar 
potential at the minimum. Moreover, by fine-tuning $D$, one 
can have a dS minimum close to zero. 
\end{enumerate}

We now employ the basic philosophy employed in KKLT model to the 
$4d$, $\mathcal{N} = 1$ supergravity to investigate the 
issue of small and positive cosmological constant with the generation of  soft 
susy breaking terms at the TeV scale. In the present context,
we restrict ourselves to the 
tree level contribution of the cosmological constant,
while it is possible to limit the effect of radiative corrections in this
computation\cite{ulrich}.
We will show how the K\"ahler correction may lead to the AdS/dS nature of the 
scalar potential.



\section{An Extended Scenario}
In this section a scenario in the context of $4d$, $\mathcal{N} = 1$ 
supergravity is constructed with an AdS minimum where 
supersymmetry is broken at an intermediate energy scale (lower than the 
Planck scale and higher than the TeV scale). In this
construction $\Phi_M$ is assumed to run over the superfields in the 
hidden sector as well as in the observable sector. 
The superpotential and K\"ahler 
potential which represent the hidden sector contributions are denoted by $\hat W$ 
and $\hat K$.  

While generating the small AdS minimum at the hidden sector, 
we will now try to address  
the following question in context of the present 
scenario vis-a-vis the  KKLT model on a very general ground.

\begin{itemize}
\item
Does the inclusion of the extra modulus in the set-up improves our understanding of the intermediate local supersymmetry breaking 
and helps to generate a vanishingly small positive cosmological constant as well as viable
susy breaking terms at the observable sector?
\end{itemize}

In seeking a probable answer to this question an additional hidden sector moduli $S$ is inserted in the superpotential 
and K\"ahler potential  (\ref{superpot4}). 
In particular, the additional term in the superpotential is intended for 
providing an mechanism for susy breaking at the intermediate scale. 
The  modified superpotential assumes the form,

\begin{equation}
\label{superpot} 
\hat W= W_0+A e^{- a \tilde T}+f(\tilde S).
\end{equation}

In the present set-up the tree level K\"ahler potential may receive corrections from 
various sources, e.g string loop 
correction (higher order in $\alpha'$ and $g_s$). With this possibility in 
mind a correction term $h(\tilde{T} + \tilde{T}^*)$ is 
appended to the tree level K\"ahler potential. 
It turns out that the correction term has a 
crucial role to play in determining the AdS nature of the potential. The 
corrected K\"ahler potential has the form,

\begin{equation}
\label{kahler2}
\hat K=M_p^2\left[-3 \log (\tilde{T} +\tilde{T}^*)-\log (\tilde{S} + 
\tilde{S}^*)+h(\tilde{T} + \tilde{T}^*)\right],
\end{equation}

Where  $\tilde S=\frac{S}{M_p}$ and $\tilde{T}=\frac{T}{M_p}$.
For these choices, the scalar potential (\ref{pot2}) can be written as,

\begin{equation}
\label{scalarpot}
V_h=\frac{1}{M_p^2}\frac{e^{ h}}{16 \tilde s \tilde t^3}\left[4  \tilde s^2 \left
[f'(\tilde s)-\frac{\hat W}{2 \tilde s} \right]^2+  \frac{1}
{(\frac{3}{4 \tilde t^2}+h'')} \left[a A e^{-a \tilde t} +\hat W 
 \left(\frac{3}{{2 \tilde t}}-h'\right)\right]^2-3 \hat W^2 \right].
\end{equation}

Here, $V_h$ stands for the scalar potential in the hidden sector only, 
$\tilde t$ and $\tilde s$ are the real components 
of the complex moduli $\tilde T$ and $\tilde S$ 
respectively. Following \cite{KKLT} we assume $W_T = W_0 + A e^{-a \tilde{t}}$, 
with the condition for susy preserving minimum (\ref{condition}),

\begin{equation}
W_T = \frac{A a e^{- a \tilde{t}_0}}{-\frac{3}{2 \tilde{t_0}} + h'(\tilde{t_0})}, 
\label{constraint1}
\end{equation}

Here $t_0$ and $s_0$ are the VEVs of the real superfield $t$ and $s$ 
respectively. 
Remembering that the SUSY breaking at the observable sector at the scale $\Lambda$ ( $\sim$ Tev ) results into
positive vacuume energy $\sim$ $\Lambda^4$, whereas the observed value is very very tiny, we arrive at the constraint
condition 

\begin{equation}
\label{const2} 
(1+\sqrt{3})\left[\frac{A a e^{-a \tilde t_0}}
{-\frac{3}{2 \tilde t_0}+h'(\tilde 
t_0)}+f(\tilde 
{s}_0)\right]=2 
\tilde s_0 f'(\tilde {s}_0).
\end{equation}
\noindent

The constraint (\ref{const2}) puts restrictions on the parameters and choice of the  functions $h$ and $f$. 
Imposing this constraint condition allows the scalar potential of the hidden sector to be written as,

\begin{equation}
\label{scalarpot2}
V_{h0}=\frac{1}{M_p^2}\frac{e^{h}}{16 \tilde s_0 \tilde t_0^3} \times \frac{1}{( h''(\tilde t_0)+
\frac{3}{4 \tilde t_0^2})}f^2(\tilde s_0) \left(\frac{3}{2 \tilde t_0}-h'(\tilde t_0)\right)^2.
\end{equation}
where $V_{h0}$ is the VEV of the scalar potential(\ref{scalarpot}).
Thus the nature of the minimum depends entirely on the choice of the function $h(\tilde t)$. 
In the following section an example has been 
presented in this regard. For a value of 
$f(\tilde{s}) \sim 10^{25} ~\rm GeV^3$ at VEV, vacuum energy
$V_{h_0} \sim -10^{12} ~\rm GeV^4$ 
can be produced provided all the dimensionless quantities are of $O(1)$. 
This, together with the contributions from the 
observable sector in the scalar potential $V_{obs} \sim 10^{12} ~\rm GeV^4$ 
may conspire to produce a small but positive cosmological constant.

The above analysis implies that the parameter $A \sim \Lambda_N^3$ should have a mass scale
with $\Lambda_N \simeq 10^{14} ~\rm GeV$ to produce viable soft breaking terms. 
This will be justified in the following sections. However admitting such large
value for $A \sim 10^{42}~\rm GeV^3$, the condition (\ref{const2}) can only be satisfied
if $f'(\tilde s)|_{VEV} \sim A >> f(\tilde s)|_{VEV}$ holds.

\section{Extremization}
In this section the issue of the stability of the 
scalar potential (\ref{scalarpot}) is addressed. 
The potential is required to satisfy the conditions for global minimum 
with well-defined VEV for all the fields.  
In this section the conditions for extremum for (\ref{scalarpot}) 
are determined  for arbitrary $f(\tilde{s})$ and $h(\tilde{t})$ where the 
constraints (\ref{constraint1}) and (\ref{const2}) have been used.

The first derivative of (\ref{scalarpot}) with respect to 
the moduli $\tilde s$ yields the condition,
\begin{equation}
\label{potmint1}
\begin{split}
\frac{\partial V_h}{\partial \tilde s}|_{\tilde s_0,\tilde t_0} =
& \frac{1}{16 \tilde s_0^2 \tilde t_0^3(3+4 \tilde t_0^2 h''[\tilde t_0])} 
[(2-\sqrt{3}) e^{h[\tilde t_0]}(4 \tilde s_0^2 f'[\tilde s_0](-\sqrt{3}f'[\tilde s_0]+(3+\sqrt{3} \tilde s_0 f''[\tilde s_0])\\
&(3+4 \tilde t_0^2 h''[\tilde t_0]) - (2+\sqrt{3})f[\tilde s_0]^2
(3+4 \tilde t_0 (h'[\tilde t_0](-3 + \tilde t_0 h'[\tilde t_0])-2 \tilde t_0 h''[\tilde t_0]))\\ 
&+2 \tilde s_0 f[\tilde s_0] (-2(2+\sqrt{3}) \tilde s_0 f''[\tilde s_0] 
(3+4 \tilde t_0^2 h''[\tilde t_0]) + f'[\tilde s_0] (12 +9 \sqrt{3} +h[\tilde t_0]\\
&+(-3(2+\sqrt{3})h'[\tilde t_0] + (2+\sqrt{3})\tilde t_0 h'[\tilde t_0]^2 -2 \tilde t_0 h'[\tilde t_0] ))))] = 0
\end{split}
\end{equation}
\noindent
Similarly the derivative with respect to the moduli field 
$\tilde t$ gives rise to the condition,

\begin{equation}
\label{potmint2}
\begin{split}
\frac{\partial V_h}{\partial \tilde t}|_{\tilde s_0,\tilde t_0} = 
&\frac{1}{16 \tilde s_0 \tilde t_0^4 (3+4 \tilde t_0^2 h''[\tilde t_0])^2}((2-\sqrt{3})e^{h[\tilde 
t_0])]}(2(3+\sqrt{3})\tilde s_0^2 f'[\tilde s_0]^2 (-3+2 \tilde t_0 h'[\tilde t_0])\\
&(3+4 \tilde t_0^2 h''[\tilde t_0])^2 + \tilde s_0 f[\tilde s_0] f'[\tilde s_0] (3+4 \tilde t_0^2 h''[\tilde t_0])
(9(11+7 \sqrt{3}+2(1+\sqrt{3})a \tilde t_0) \\
&+2 \tilde t_0 (2(1+\sqrt{3})\tilde t_0 (9+2 a \tilde t_0)h'[\tilde t_0]^2 
- 4(1+\sqrt{3})\tilde t_0^2 h'[\tilde t_0]^3 + 6 (9+5 \sqrt{3})\tilde t_0 h''[\tilde t_0] \\
&-h'[\tilde t_0](3(11+9 \sqrt{3}+4(1+\sqrt{3})a \tilde t_0) + 4(3+\sqrt{3}\tilde t_0^2 h''[\tilde t_0])))\\
&+ (2+\sqrt{3})f[\tilde s_0]^2 (-27 +\tilde t_0 (4 \tilde t_0^2 h'[\tilde t_0]^3 (3+4 \tilde t_0^2 h''[\tilde t_0]) \\
&+12 \tilde t_0 (-6 h''[\tilde t_0]+4 \tilde t_0^2 h''[\tilde t_0]^2-3 \tilde t_0 h''[\tilde t_0])\\
&-16 \tilde t_0 h'[\tilde t_0]^2 (3 +6 \tilde t_0^2 h''[\tilde t_0]+ \tilde t_0^3 h'''[\tilde t_0])
+ 3 h'[\tilde t_0](27+68 \tilde t_0^2 h''[\tilde t_0] + 16 \tilde t_0^3 h'''[\tilde t_0]))))).
\end{split}
\end{equation}
Results of the calculation of the second derivatives
for suitable choices of $f(\tilde{s})$ and $h(\tilde{t})$ indicates that 
the conditions stated above can be used to yield a stable global minimum 
for the scalar potential without having any flat direction. 
In section (5) with a specific choice 
of $f(\tilde{s})$ and $h(\tilde{t})$, the existence of a global 
AdS minimum for the potential will be  illustrated.

\section{Soft susy breaking parameters }

It is worthwhile to discuss the low energy phenomenology encoded in the soft susy breaking parameters.
We consider the chiral superfield of $\mathcal{N}=1$ supergravity
(see section 1.1) $\Phi_{M} ={h_m,C^\alpha}$ where $h_m$ 
corresponds to heavily massive components and constitute the hidden sector and $C^\alpha$ are lighter mass components
and constitute the observable sector. While the Latin indices run over the entire set of hidden sector fields, 
the Greek indices denotes the set of the MSSM fields. The superpotential and K\"ahler potential can be 
Taylor expanded with respect to the observable sector fields as (for details see \cite{ibanez}). 
\begin{equation}
\label{superpotential}
W=\hat{W}(h_m)+\frac{1}{2} \mu_{\alpha \beta}(h_m) C^{\alpha} C^{\beta} +\frac{1}{6} Y_{\alpha \beta \gamma}(h_m) C^{\alpha} 
C^{\beta}C^{\gamma}+.....
\end{equation}
and 
\begin{equation}
\label{kahler}
K=\hat{K}(h_m,h_m^*)+\tilde K_{\bar \alpha \beta}(h_m,h_m^*){C^*}^{\bar \alpha} C^{\beta}+ 
\frac{1}{2}(Z_{\alpha \beta} C^{\alpha} C^{\beta} + h.c) + ...
\end{equation}
As already mentioned the $\hat W$ and $\hat K$ represent the hidden sector components 
of the superpotential and K\"ahler potential. Similarly, 
in (\ref{kahler}) the observable sector K\"ahler metric is given by $\tilde{K}_{\alpha \bar{\beta}}$.  
The above expansions for the superpotential and K\"ahler potential can 
be plugged in the generic expression for the scalar potential (\ref{pot1}) giving rise to the full scalar potential of 
the theory.
Choosing the K\"ahler metric of the observable sector in diagonal form  as

\begin{equation}
\label{delta}
{\tilde K}_{\overline{\alpha} \beta}(h_m,h_m^*) = 
{\delta}_{\overline {\alpha} \beta} 
{\tilde K}_{\alpha}(h_m,h_m^*) 
\end{equation}

and canonically normalizing the fields we get the scalar part 
of the the effective soft Lagrangian 

\begin{equation}
\label{softLagrangian}
{\cal L}_{soft} = 
- m_{\alpha}^2 \tilde{C}^{*\overline {\alpha}}
\tilde{C}^{\alpha} - \left(\frac{1}{6} A_{\alpha \beta \gamma} Y_{\alpha \beta \gamma} \tilde{C}^{\alpha} \tilde{C}^{\beta} \tilde{C}^{\gamma}
+ B {\mu} H_1 H_2+h.c.\right) 
\end{equation}

where $\tilde{C}^\alpha$ is the canonically normalized MSSM fields, $m_\alpha$ is the scalar mass, $A_{\alpha \beta \gamma}$ 
is the trilinear coupling and $B$ is the bilinear Higgs coupling. Here
$H_1,H_2$ are the two Higgs fields in the MSSM. 
These three soft mass parameters can be expressed 
in terms of the K\"ahler potential and superpotential as follows. 
  
\begin{gather}
\label{mmmatrix}
{m}_{\alpha}^2 = 
\left(m_{3/2}^2+\frac{V_{h0}}{M_P^2}\right) - {\overline{F}}^{\overline{m}} F^n 
\partial_{\overline{m}}\partial_n \log{\tilde K_{\alpha}},\\
\label{mmmatrix2}
A_{\alpha\beta\gamma} = 
F^m \left[{\hat K}_m/{M_P^2} + \partial_m \log Y_{\alpha\beta\gamma} 
- \partial_m \log({\tilde K_{\alpha}} {\tilde K_{\beta}}
{\tilde K_{\gamma}}) \right],\\ 
\label{mmmatrix3}
\begin{split}
B = &{\hat \mu}^{-1}({\tilde K}_{H_1}{\tilde K}_{H_2})^{-1/2}
\left\{ \frac{ {\hat W}^*}{|{\hat W}|} e^{{\hat K}/2} \mu 
\left( F^m \left[ {\hat K}_m + \partial_m \log\mu\right.\right.\right.
\left.\left.-\ \partial_m \log({\tilde K_{H_1}}{\tilde K_{H_2}})\right]
- m_{3/2} \right)
\\ 
&+ 
\left( 2m_{3/2}^2+V_0 \right) {Z} -
m_{3/2} {\overline{F}}^{\overline{m}} \partial_{\overline{m}} Z
+\ m_{3/2} F^m \left[ \partial_m Z - 
Z \partial_m \log({\tilde K_{H_1}}{\tilde K_{H_2}})\right]
\\ 
&\left.-\ {\overline{F}}^{\overline{m}} F^n 
\left[ \partial_{\overline{m}} \partial_n Z - 
 \partial_{\overline{m}} Z 
\partial_n \log({\tilde K_{H_1}}{\tilde K_{H_2}})
\right] \right\}.  
\end{split}
\end{gather}


While calculating the bilinear coupling parameter, we assume $Z_{\alpha\beta}
= 0$\cite{ibanez} in the K\"ahler potential (\ref{kahler}) for
simplicity. 
In this limit, one may use $\hat{\mu} = 
\left(  \frac{ {\hat W}^*}{|{\hat W}|} e^{\hat K/2} {\mu}\right)
 ({\tilde K}_{H_1}{\tilde K}_{H_2})^{-1/2}$,
where $\tilde{K}_{H_1}$ and $\tilde{K}_{H_2}$ 
are the K\"ahler metric coupled to the Higgs fields $H_1$ and $H_2$ respectively. In all the above 
three expressions, ((\ref{mmmatrix}) - (\ref{mmmatrix3})), $F_m$ is the hidden sector auxiliary field and $\widehat{K}_m$ 
is the derivative of the K\"ahler potential with respect to the hidden sector field $h_m$.   
Now using $h_m = \{T,S\}$ and  
$\tilde{K}_{\alpha} = (\tilde T + \tilde T^*)^{-1}$,
the various soft parameters can be computed in the present case. Using   
the expressions for the superpotential(\ref{superpot}), the K\"ahler potential(\ref{kahler2}), 
the constraint conditions (\ref{constraint1}), (\ref{const2}) and also 
the conditions 
for extremum ((\ref{potmint1}) and (\ref{potmint2})) for the 
scalar potential (\ref{scalarpot}) the VEV for all
moduli fields may be fixed. Though the above expressions(\ref{mmmatrix},~
\ref{mmmatrix2},~\ref{mmmatrix3}) have been 
expressed in the units $M_p=1$ unit , to get the correct order of magnitude we now calculate 
the various soft terms bringing $M_p$ back in the expressions. 
Then the gravitino mass $m_{3/2}$ and scalar mass term 
$m_{\alpha}$ are given by

\begin{gather}
\label{gravitino}
m_{3/2}^2=\frac{1}{M_p^4} e^{ h(\tilde t_0)} \frac{1}{16 \tilde s_0 \tilde t_0^3} \left[f(\tilde s_0)
+\frac{A a e^{-a \tilde t_0}}{-\frac{3}{2 \tilde t_0}+h'(\tilde t_0)}\right]^2\\ 
\label{softmass}
\begin{split}
m_{\alpha}^2 = ~   & m_{3/2}^2 \\
&+ \frac{1}{M_p^4}\frac {e^{h(\tilde t_0)}}{{16 \tilde s_0 \tilde t_0^3}}  \frac{1}{\left(\frac{3}{4 \tilde t_0^2}
+h''(\tilde t_0) \right)} \left[f^2 ( \tilde s_0) \left(\frac{3}{2 \tilde t_0}-h'(\tilde t_0) \right)^2 \right]\left[1-
\frac{1}{4 \tilde t^2_0} \frac{1}{\left(\frac{3}{4 \tilde t_0^2}+ h''(\tilde t_0)\right)}\right]
\end{split}
\end{gather}
\noindent

Considering the VEV for all moduli fields $\sim M_p$ and $h''(\tilde t_0)$
as well as $a$ are very close to $O(1)$, the gravitino mass $m_{3/2}$ may 
become 
$\sim$ TeV. With $A \sim \Lambda_N^3 
\sim 10^{42} ~\rm GeV^3$ gravitino mass $\sim$ TeV 
can be generated. It is important to note that, the contribution arising from $F^T$ 
would involve only $f(\tilde s_0)$ at the minimum. Then, accepting the
fact that $f(\tilde s_0)$ assumes a value which can produce a small
vacuum energy, $F^T$ cannot provide any significant contributions towards the
soft breaking parameters. Thus in Eq.(\ref{softmass}),
the scalar mass term $m^2_\alpha$ parameter includes the $m^2_{3/2}$ as 
the leading
term.
Neglecting the terms associated with $f(\tilde s_0)$, 
the trilinear term can be calculated as

\begin{equation}
\label{trilinear}
\begin{split}
&A_{\alpha \beta \gamma} \simeq 
\frac{1}{M_p^2}~\frac{e^{h(\tilde t_0)/2}}{\tilde s_0^{1/2} \tilde t_0^{3/2}}~ s_0^{2}
~\left(f'(\tilde s_0) -\frac{\hat W}{2 \tilde s_0}\right)\left[-\frac{1}{2 \tilde{s}_0}
+\left(\partial_{\tilde{s}} \log Y_{\alpha \beta \gamma}\right)|_{VEV}\right].
\end{split}
\end{equation}
\noindent

Using the constraint (\ref{const2}) and the 
expression for the gravitino mass 
(\ref{gravitino}) this simplifies to, 
\begin{equation}
\label{trilinear2}
A_{\alpha \beta \gamma} \simeq  m_{3/2} \tilde{s}_0\left[-\frac{1}{2 \tilde{s}_0}
+\left(\partial_{\tilde{s}} \log Y_{\alpha \beta \gamma}\right)|_{VEV}\right].
\end{equation}
\noindent
Similarly the bilinear coupling reduces to, 

\begin{equation}
\label{bilinear}
B \simeq m_{3/2} \tilde{s}_0 \left[-\frac{1}{2 \tilde{s}_0} + \partial_{\tilde{s}} \log \mu|_{VEV} \right]-m_{3/2}. 
\end{equation}
\noindent

It can therefore be concluded that the trilinear (\ref{trilinear2}) 
and bilinear parameters (\ref{bilinear})
are also $\sim \rm TeV$. Thus following the discussion in sections
(2),(3) and (4) it is easy to infer that the inclusion of an additional moduli
field which breaks the susy may produce small vacuum
energy in the hidden sector. The correction term is crucial to determine the
nature of the minimum. The susy breaking parameters at the observable sector 
have the TeV scale values. In the following section an example in support of the above proposal is presented.

\section{A toy model}

We construct a specific model to illustrate the general scenario 
discussed so far. 
It has already been mentioned that the condition  
$(h''(\tilde t_0) + \frac{3}{4 \tilde t_0^2}) <0$ 
in (\ref{scalarpot2}) leads to an AdS minimum.
The correction term in the 
K\"ahler potential is chosen as 

\begin{equation}
\label{Kahlercorrect}
h(\tilde T + \tilde T^*) \equiv 
-\frac{\alpha}{(\tilde T+\tilde T^*)} - \alpha_1 (\tilde T + \tilde T^*)^2,
\end{equation}

where $\alpha$ and $\alpha_1$ are two independent parameters. It is noteworthy
that in (\ref{Kahlercorrect}) appropriate choice of parameters $\alpha$ and $\alpha_1$ may produce the dS vacuum
as well. However in this particular context
only the nature of AdS vacua is studied.

The function $f(\tilde{s})$ is chosen to be,
\begin{equation}
\label{fs}
f(\tilde{s}) = m^3((\ln{\tilde{s}})^2 + k \ln{\tilde{s}} + k_1 \tilde{s}),
\end{equation}

where $k$ and $k_1$ are free parameters and $m$ is the scale of
the potential. For this choice the superpotential 
corresponding to the moduli field $T$ at the VEV becomes,

\begin{equation}
\label{superpotVEV}
W^{T}=\frac{A a e^{-a \tilde t_0}}{-\frac{3}{2 \tilde t_0}+\frac{\alpha}{4 \tilde t_0 ^2 } - 4\alpha_1 \tilde{t}_0}.
\end{equation}
Furthermore the constraint (\ref{const2}) acquires the form,

\begin{equation}
\begin{split}
\label{const3}
&(1+\sqrt{3})\left[\frac{A a e^{-a \tilde t_0}}{-\frac{3}{2 \tilde t_0}+\frac{\alpha}{4 \tilde t_0 ^2 } - 
4\alpha_1 \tilde{t}_0 }+ m^3 ((\ln{\tilde{s}_0})^2 + k \ln{\tilde{s}_0} + k_1 \tilde{s}_0) 
\right]\\
&= 2\tilde s_0 m^3 \left(\frac {2 \ln{\tilde{s}_0}} {\tilde s_0} + \frac{k}{\tilde s_0}
+ k_1\right) . 
\end{split}
\end{equation}

\begin{figure}[!h]
\centering
\includegraphics{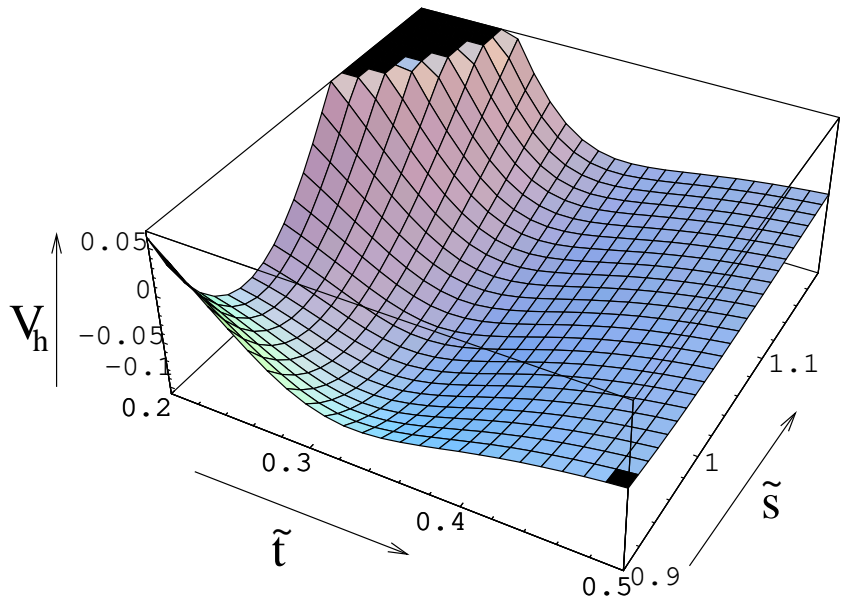}
\caption{3D plot of $V_{h}$ vs moduli $\tilde s$ $\&$ $\tilde t$ in units of $10^{12}~\rm GeV^4$
for parameter values $A = - 1.60, W_0 = 1.7, a = 0.1, k= 0.1, \alpha = 0.72, \alpha_1 = 1.5$. 
Here $A$ and $W_0$ represent values in the unit of $\Lambda_N^3$.}
\label{3dminimum}
\end{figure}

\noindent
The VEV of the scalar potential for the hidden sector $V_h$ now turns out to be,
\begin{equation}
\label{pothidden}
V_{h0} = \frac{1}{M_p^2} e^{\frac{-\alpha}{2 \tilde t_0} - 
4 \alpha_1 {{\tilde t}^2}_0}\frac{m^6}{4 \tilde s_0 \tilde t_0} \frac{((\ln{\tilde{s}_0})^2 
+ k \ln{\tilde{s}_0} + k_1 \tilde{s}_0)^2}{\left(3-\frac{\alpha}{\tilde t_0} 
-8 \alpha_1 {{\tilde t}^2}_0\right)} \left(\frac{3}{2 \tilde t_0}
-{\frac{\alpha}{\tilde 4 t_0^2} + 4 \alpha_1 \tilde t_0}\right)^2.
\end{equation}
From the preceding equation it can be estimated that with $k_1 \sim
10^{-17}$ and $m \sim 10^{14} ~\rm GeV$ the vacuum energy becomes 
$\sim 10^{12} ~\rm GeV^4$, provided the minimum of the potential 
corresponds to $\tilde s_0=1$.
Such a minimum can indeed be found for suitable values of the parameters which
is depicted in Fig(\ref{3dminimum}). 

While $\alpha$ and $\alpha_1$ are chosen so as to 
have the AdS minimum, the constant parameter $k$ is  
set at $O(1)$ values to satisfy the condition (\ref{const3}). All the soft susy breaking 
parameters are now estimated assuming $k_1 \sim
10^{-17}$ and $m \sim 10^{14} ~\rm GeV$ along with K\"ahler correction 
as given via Eq.\ref{Kahlercorrect}. 
Then the gravitino mass given by (\ref{gravitino}) becomes,

\begin{equation}
\label{gravitino3}
m_{3/2}^2=\frac{1}{M_p^4} e^{-\frac{\alpha}{2 \tilde t_0} - 
4 \alpha_1 {{\tilde t}^2}_0}~ \frac{m^6}{16 \tilde s_0 \tilde t_0^3}\left[\frac{1}{(1+\sqrt{3})} ~
2 \tilde s_0 (\frac{k}{\tilde{s}_0} + 2\frac{\ln{\tilde{s}_0}}{\tilde{s}_0})\right]^2.
\end{equation}

In (\ref{gravitino3}), the derivative  
$f'(\tilde{s}_0) \simeq (\frac{k}{\tilde{s}_0} 
+ 2\frac{\ln{\tilde{s}_0}}{\tilde{s}_0})$ have been used ignoring terms with 
$k_1 \sim 10^{-17}$.
The soft scalar mass (\ref{softmass}) can be recast as,
\begin{equation}
\label{massparmtr}
\begin{split}
m_{\alpha}^2= ~ &m_{3/2}^2  +\frac{1}{M_p^4} \frac{e^{- \frac{\alpha}{2 \tilde t_0} 
- 4 \alpha_1 {{\tilde t}^2}_0}}{16  \tilde s_0 \tilde t_0^3}  \frac{m^6}{\left(\frac{3}{4 \tilde t_0^2}
-\frac{\alpha}{4\tilde t_0^3}- 2\alpha_1 \right)}\\
& \left[((\ln{\tilde{s}_0})^2 + k \ln{\tilde{s}_0} + k_1 \tilde{s}_0) \left(\frac{3}{2 \tilde t_0}
- \frac{\alpha}{4 \tilde t_0^2} + 4 \alpha_1 
\tilde{t}_0 \right)\right]^2 
\left[1-\frac{1}{4 \tilde t_0^2} \frac{1}{\left(\frac{3}{4 \tilde t_0^2}
-\frac{\alpha}{4 \tilde t_0^3 } -2 \alpha_1 \right)}\right].
\end{split}
\end{equation}

The trilinear and bilinear couplings can now be calculated as before. 
The leading contributions to the 
trilinear coupling (\ref{trilinear2}) and bilinear couplings (\ref{bilinear}) come from the gravitino mass $m_{3/2}$,
which can be estimated from (\ref{gravitino3}). 

It is worthwhile is see that whether such a scenario could actually predict
a global minimum with $\tilde s_0 =1$. To investigate that, we 
illustrate our result with  a $3D$-plot (fig \ref{3dminimum}) of the hidden sector scalar potential. 
It clearly exhibits the existence of a global AdS 
minimum of the potential which is essential for its stability. 
It can be seen that the resulting cosmological constant 
has been very finely tuned to make it compatible with the near Minkowskian but de- Sitter nature of the space-time  
of the present day universe. 
It is however worth pointing out here that the existence of the 
global minimum is 
extremely sensitive to the values of the parameters $\alpha, \alpha_1, A, {W}_0, a, k$ and $k_1$, thus rendering the 
model as fine tuned.

\section{Discussions}

String theory admits of numerous vacua, but only a very few of them are compatible with phenomenologically viable models in the 
low energy limit. It is therefore worthwhile to investigate the possible features of string inspired supergravity models from 
both particle phenomenology as well as cosmological point of view. This work is an attempt to establish the required constraints 
on the structures of these models in order to get the desired features. In this work we have considered a 
general supergravity scenario which simultaneously may address the issues of early inflationary 
phase of the universe, smallness of the 
observed cosmological constant in the present epoch as well as the TeV scale supersymmetry
breaking which are known to be deeply connected to each other.
In our analysis we assume the existence of a hidden sector with appropriate VEV of the moduli fields which couples gravitationally
with the observable sector.Considering a
dual scenario namely flux compactification and gaugino condensation at two different scales, we achieve to bring  out the 
necessary constraints on the Kahler potential and superpotential so that the  desired phenomenological values of the 
observable sector parameters can be obtained with no flat direction in the
resulting scalar sector.We have shown that at least two hidden sector fields are necessary to achieve this.

We conclude by presenting a simple model in support of the analysis carried out in this work to 
illustrate that all the generic constraints 
on the K\"ahler and superpotential may be mutually compatible. Various parameters of the observable sector are 
estimated for this model. Our constraints 
thus put a set of restrictions on the class of supergravity models which they have to satisfy to yield phenomenologically 
viable results so that the following features can be achieved concomitantly :
\begin{itemize}
\item Soft supersymmetry breaking parameters are generated in the observable sector at the TeV scale.
\item All the flat directions in the hidden sector are removed to have well-defined estimates of the soft breaking parameters.
\item The resulting vacuum energy appearing from various mechanisms add up to a tiny positive value consistent 
with the presently observed de-Sitter character of the 
Universe. 
\end{itemize}

\section{Acknowledgment}
DD thanks P2I, CNRS for the support received as a post-doctoral fellow.
The authors would like to thank Pradipta Ghosh , Koushik Ray and Sourov Roy of the Department of Theoretical Physics, 
IACS, Kolkata, for fruitful discussions and suggestions.

\end{document}